\documentclass[aps,prl,twocolumn,superscriptaddress,10pt,floatfix]{revtex4-2}

\usepackage{amsmath, amssymb}
\usepackage{mathrsfs}
\usepackage{graphicx}
\graphicspath{{Figs/}}
\usepackage{bm}
\usepackage{braket}
\usepackage{multirow}
\usepackage[colorlinks=true, citecolor=blue]{hyperref}

\newcommand{\diff}{\mathrm{d}}
\newcommand{\imag}{\mathrm{Im}\,}
\newcommand{\real}{\mathrm{Re}\,}

\newcommand{\epn}{\mathrm{e}}
\newcommand{\al}{\alpha}
\newcommand{\be}{\beta}

\newcommand{\om}{\omega}

\newcommand{\lam}{\lambda}

\newcommand{\nt}{\notag \\}

\newcommand{\mcal}[1]{\mathcal{#1}}

\begin{document}

\title{
Anomalous inverse Faraday effect for graphene quantum dots in optical vortices
}

\author{Zi-Yang Xu}
\affiliation{College of Physics, Chongqing University, Chongqing 401331, China}

\author{Wei E. I. Sha}
\affiliation{College of Information Science and Electronic Engineering, Zhejiang University, Hangzhou 310027, China}

\author{Hang Xie}

\email[Corresponding author: ]{xiehang2001@hotmail.com}

\affiliation{College of Physics, Chongqing University, Chongqing 401331, China}
\affiliation{Chongqing Key Laboratory for Strongly-Coupled Physics, Chongqing University, Chongqing 401331, China}

\date{\today}

\begin{abstract}

Chiral photon interactions with two-dimensional (2D) materials enable unprecedented control of quantum phenomena. 
In this paper, we report anomalous inverse Faraday effects (IFE) in graphene quantum dots (GQDs) under linearly polarized optical vortex illumination, where transferred orbital angular momentum (OAM) generates light-induced magnetic moments.
Employing our recently developed time-dependent quantum perturbation framework [Phys. Rev. B 110, 085425 (2024)], we demonstrate a counterintuitive observation: some reversed magnetic moments at off-axis positions occur—manifested as counter-rotating currents to the vortex helical wavefront. 
Phase-difference analysis and eigenmode decomposition resolve this anomaly, revealing that the OAM transfer efficiency is orders of magnitude weaker than its spin counterpart.
This work establishes a new paradigm for optical OAM-to-magnetization conversion in quantum-engineered 2D systems. 

\end{abstract}

\maketitle

\textit{Introduction}---
Optical vortices, characterized by helical wavefronts and phase singularities, carry quantized orbital angular momentum (OAM) and spin angular momentum (SAM, via selective circular polarization) \cite{Allen1992Orbital, Orlov2002Propagation}. 
These structured beams enable unprecedented control over condensed matter systems \cite{QuinteiroRosen2022Interplay}. 
Their interactions with two-dimensional (2D) materials generate topologically driven currents, such as orbital photogalvanic effects in graphene/WTe$_2$ \cite{Ji2020Photocurrent, Han2025Topologically} and quantum Hall-like transport in annular devices \cite{Session2025Optical}. 
While vortex-induced phenomena have been extensively explored across diverse systems—including atomic/molecular ensembles \cite{Schmiegelow2016Transfer, Lembessis2013Enhanced, Babiker2019Atoms, Bougouffa2020Quadrupole, Babiker2002Orbital, Alexandrescu2006Mechanism}, quantum dots \cite{Fong2018Scheme, Quinteiro2009Electronic, Mahdavi2020Manipulation}, and topological insulators \cite{Shintani2016Spin}—the transfer of photonic angular momentum to electronic or lattice systems remains a complex and contentious subject, particularly in mesoscopic regimes. 
Typically, this transfer generates a photoinduced magnetic moment in the material—a phenomenon termed the inverse Faraday effect (IFE) \cite{Ziel1965Optically-Induced}. 
Of particular interest is the manifestation of IFE in quantum-confined systems driven by optical vortices, which represents an uncharted frontier \cite{Karakhanyan2022Inverse, Mahdavi2020Manipulation}.

Graphene quantum dots (GQDs), with their tunable electronic confinement and symmetry-governed optical selection rules, provide an ideal platform for probing OAM/SAM conversion \cite{Yan2019Recent, Kavousanaki2015Optically, Pohle2018Symmetry}. 
While our recent work analyzed IFE in GQDs driven by conventional circularly polarized light (CPL) \cite{Xu2024Photoinduced}, the spatially structured vortex field introduces new control dimensions. 
Notably, although vortex-driven OAM transfer is dipole-forbidden for atomic electrons \cite{Babiker2002Orbital}, studies demonstrate this constraint is lifted in molecular/semiconductor nanostructures \cite{Alexandrescu2006Mechanism, Quinteiro2009Electronic}.

Here, we report a spatially switched IFE in GQDs driven by vortex beams. 
Employing a novel time-dependent quantum perturbation framework, we demonstrate that the GQD exhibits a \textit{reversed light-induced magnetic moment}—opposite to the OAM direction of the incident vortex—when positioned at specific field regions. 
This anomalous magnetization admits rigorous self-consistent interpretation through phase-difference analysis and eigenmode decomposition.
Furthermore, we reveal the angular momentum conversion ratio in circularly polarized vortex light, establishing SAM-dominant transfer over OAM. 
Our work establishes a new paradigm for optically controlled nanomagnetism in quantum-confined 2D nanostructures.

\textit{Model}---
The $N$-atom GQD occupies the $xy$-plane, irradiated by $z$-propagating optical vortices. 
Linear polarization nullifies photon spin effects (circular case in Appendix B). 
Using a semiclassical description where optical vortex fields obey Maxwell's equations, we express the Hamiltonian as
\begin{align}
\label{eq:H}
        H = H _ { 0 } + H _ { i n t } . 
\end{align}
The first term ($H_0$) is the unperturbed Hamiltonian of GQD, written in the tight-binding (TB) model as $H_{0} =-t\sum_{\left \langle i,j \right \rangle } \hat{c}_{i}^{\dagger}\hat{c}_{j}$ \cite{Kavousanaki2015Optically, Xu2024Photoinduced, Wallace1947TheBand},
where $\hat{c}_{i}^{\dagger}$ and $\hat{c}_{i}$ are the creation and annihilation operators for an electron at the atomic orbital with index $i$; 
in the summation, $\left \langle i,j \right \rangle$ denotes nearest-neighbor pairs $i$ and $j$; $t$ is the hopping energy which is $t=2.7$\,eV for GQD \cite{Reich2002Tightbinding, CastroNeto2009Theelectronic}.

The second term ($H_{int}$) is the interaction Hamiltonian, which can be written as a multipolar series expansion in terms of the appropriate coordinate $\bm R$ within the GQD \cite{Lembessis2013Enhanced, ZuritaSanchez2002Multipolar, Quinteiro2015Formulation, Rossi2002Theory}.
\begin{align}
\label{eqn:H_int}
        H _ { i n t } = H _ { d p } + H _ { q p } + \cdots ,
\end{align}
where $H_{dp}$ is the electric dipole interaction between the GQD and the vortex field. Explicitly we have
\begin{align}
\label{eqn:H_dp}
        H _ { d p } = - \bm d \cdot \bm E (\bm R , t ) ,
\end{align}
where $\bm d=q \bf x$ with $\bf x$ $ = (x,y)$ as the internal position vector, is the electric dipole moment vector, $q$ the electron charge; $\bm E (\bm R , t )$ the electric field (E-field) vector. 
And $H_{qp}$ in Eq.~(\ref{eqn:H_int}) denotes the electric quadrupole interaction, which can be neglected compared to the dipole moment at locations away from phase singularity [here defined as $r\gtrsim{10}^{-2}\lam$ (wavelength)]; see Supplemental Material (SM) for details \cite{Supplementary}.

For a generic linearly polarized vortex beam with wave vector $k$, the E-field in cylindrical coordinates $\bm R=(r,\theta,z)$ is \cite{Romero2002Aquantum, Babiker2002Orbital}
\begin{align}
\label{eqn:E(Rt)}
        \bm E  ( \bm  R  , t ) = \hat { \bm i } E ( r , z ) \exp ( i l \theta ) \exp [ i ( k z - \om t ) ] ,
\end{align}
where $\hat { \bm i }$ is the unit polarization vector, $\om$ the angular frequency, and $E ( r , z )$ the normalized radial distribution function.
Accounting for the 2D nature of GQDs, the Laguerre-Gaussian (LG) mode \cite{QuinteiroRosen2022Interplay, Romero2002Aquantum} assumes a simplified $z$-independent form
\begin{align}
\label{eqn:E(r)}
        E ( r ) = \frac { C _ { p } ^ { | l | } } { w _ { 0 } } \Big ( \frac { r \sqrt { 2 } } { w _ { 0 } } \Big ) ^ { | l | }  L _ { p } ^ { | l | }  \Big ( \frac { 2 r ^ { 2 } } { w _ { 0 } ^ { 2 } } \Big ) \exp { \Big ( \frac { - r ^ { 2 } } { w _ { 0 } ^ { 2 } } \Big ) } , 
\end{align}
where $C _ { p } ^ { | l | }=\sqrt {  2 p ! / [ \pi ( p + | l | ) ! ] } $, $w_0$ is the beam waist radius.
$L _ { p } ^ { | l | }$ is the associated Laguerre polynomial, topological charge $l$ is the azimuthal index giving an OAM of $l\hbar$ per photon, and $p$ the number of radial nodes. 
With the polarization along $x$, $H_{dp}$ in Eq.~(\ref{eqn:H_dp}) can be rewritten as $H_{dp} = -q x E ( r ) \cos  ( l \theta - \om t ).$

To model the GQD's electronic dynamics under vortex illumination, we employ time-dependent perturbation theory with $H _ { i n t }$ as perturbation. 
The wavefunction evolves from initial eigenstate $\Psi_k$ to excited states as \cite{Xu2024Photoinduced, DavidJ2005Introduction}
\begin{align}
        \Psi ( t ) &= \Psi _ { k } \epn^ { - i \omega _ { k } t } + \sum _ { k _ { 1 } } S _ { k _ { 1 } } ( t ) \bra{ \Psi _ { k _ { 1 } } } x E ( r ) \epn ^ { i l \theta } \ket{\Psi _ { k }} \nt
        &\;\;\;\; \label{eqn:Psi(t)}
        \times \Psi _ { k _ { 1 } } \epn ^ { - i \omega _ { k _ { 1 } } t } ,
\end{align}
with $S_{k_{1}}\left( t \right) =\frac{q(\epn^{i\Delta\omega t}-1)}{2\hbar\Delta\omega}$,
where $\omega _ { k }=\frac{E_k}{\hbar}$, $E_k$ is the eigenenergy of the corresponding state $\Psi_k$, $k\in1,\ldots,N$.
In the formula of time-dependent coefficient $S_{k_{1}}\left( t \right)$, $\Delta\omega=\omega_{k_1k}-\omega$, and $\omega_{k_1k}=\omega_{k_1}-\omega_k$.
We here employ the rotating wave approximation and only consider the first-order correction of perturbation.

Under the TB model ($\ket{\Psi_k}=\sum_{n}a_n^k\ket{n}$), we may transform the inner product in Eq.~(\ref{eqn:Psi(t)}) into the atomic orbital basis as $\left( x E ( r ) \epn ^ { i l \theta } \right) _ { n , n ^ { \prime } } = \bra{n} x \ket{n ^ { \prime } } E ( r _ { n ^ { \prime } } ) \epn ^ { i l \theta _ { n ^ { \prime } } }$,
where $\ket{n}$ ($\ket{n ^ { \prime } }$) represents the atomic basis, $r_{n^\prime}$ ($\theta_{n^\prime}$) is the coordinate component of the corresponding atomic position $\bm{R}_{n^\prime}$. 
This dipole matrix element originates from the multipole expansion of the atomic position.

\textit{Photoinduced currents}---
The electron current density in the quantum theory holds $\bm j  = \frac { q \hbar } { m } \imag [ \Psi ^ { \ast } ( t ) \bm { \nabla } \Psi ( t ) ]$ \cite{Xu2024Photoinduced, DavidJ2005Introduction},
where $\bm{j}$ and $m$ are, respectively, the current density vector and the electron mass. 
So we obtain three types of current densities as $\bm{j}_{\rm{type\mbox{-}\!I}}=\frac{q\hbar}{m} \imag \left (\Psi_{k}^{\ast} \bm{\nabla} \Psi_{k}\right )$,
$$  
\bm{j}_{\rm{type\mbox{-}\!II}}=\frac{q\hbar}{m} \imag \sum_{k_{1}} \left ( G_{k_{1}} \Psi_{k}^{\ast}\bm{\nabla}\Psi_{k_{1}}+G_{k_{1}}^{\ast} \Psi_{k_{1}}^{\ast}\bm{\nabla}\Psi_{k} \right ),
$$
$$
\bm{j}_{\rm{type\mbox{-}\!III}}=\frac{q\hbar}{m} \imag \sum_{k_{1} k_{2}}G_{k_{1}}^{\ast}G_{k_{2}} \Psi_{k_{1}}^{\ast}\bm{\nabla}\Psi_{k_{2}},
$$  
where $G_{k_{1}}=S _ { k _ { 1 } } ( t ) \bra{ \Psi _ { k _ { 1 } } } \al+i \be \ket{\Psi _ { k }}\epn ^ { - i \omega _ { k _ { 1 } k} t }$, $\al=x E ( r )\cos l \theta$, $\be=x E ( r )\sin l \theta$.
The type-I eigencurrent is zero in our electro-optic interaction system, confirming no ground-state photoinduced current without optical perturbation.

For light absorption, Fermi's golden rule requires $\Delta\omega \to 0$, giving $\omega _{k_{1(2)}k} = \omega$.
Considering long-time behavior, we expand $\ket{\Psi_k}$ in the atomic basis $\ket{n}$, then integrate $\bm{j}_{\rm{type\mbox{-}\!II}}$ and $\bm{j}_{\rm{type\mbox{-}\!III}}$ spatially to obtain the local currents at each atomic site $n$:
\begin{subequations}
\label{eqn:J_typeII_typeIII}
\begin{align}
        \bm{J}_{n}^{\rm{type\mbox{-}\!II}}
        &=\frac{q^{2}\pi}{m}\sum_{\left \langle n^\prime \right \rangle}\sum_{k_{1}}
        \left ( \al_{k_{1}k}\cos \omega t+\be_{k_{1}k}\sin \omega t \right ) \nt
\label{eqn:J_typeII}
        &\;\;\;\;
        \times\left (a_{n}^{k}a_{n^\prime}^{k_{1}}-a_{n}^{k_{1}}a_{n^\prime}^{k}\right )
        \bra{n} \bm{\nabla} \ket{n^\prime} \delta \left ( \Delta \omega \right ),\\
        \bm{J}_{n}^{\rm{type\mbox{-}\!III}}
        &=\frac{q^{3}\pi}{2m\hbar}\sum_{\left \langle n^\prime \right \rangle}\sum_{k_{1}k_{2}}
        \left ( \al_{k_{1}k}\be_{k_{2}k}-\al_{k_{2}k}\be_{k_{1}k} \right ) \nt
\label{eqn:J_typeIII}
        &\;\;\;\;    
        \times a_{n}^{k_{1}}a_{n^\prime}^{k_{2}}
        \bra{n} \bm{\nabla} \ket{n^\prime} t \delta \left ( \Delta \omega \right ),
\end{align}
\end{subequations}
where $\al_{k_{1}k}=\bra{ \Psi _ { k _ { 1 } } } \al \ket{\Psi _ { k }}$. 
$\left \langle n^\prime \right \rangle$ in the summation above means the site $n^\prime$ is the nearest neighbor to site $n$. 
The full derivation follows our prior work on CPL-GQD interactions \cite{Xu2024Photoinduced}, with vortex beam coordinates now denoted $(\al,\be)$ instead of $(x, y)$.
Equation~(\ref{eqn:J_typeII_typeIII}) reveals distinct current behaviors: type-II oscillates periodically while type-III grows linearly in time. 
Crucially, type-III emerges exclusively in degenerate excited states.

Characterizing the GQD's vortex-driven absorption spectrum is essential to establish current-generation conditions. 
We derive this spectrum by adapting CPL methodology to vortex illumination.
The Kubo formula for CPL spectrum \cite{Pohle2018Symmetry, Xu2024Photoinduced} derives from the Joule heating-absorbance correspondence. 
For vortex illumination, the total absorption requires integration of the heat density over the GQD area $A$:
\begin{align}
\label{eqn:joule}
        Q _ { j o u l e } = \frac { 1 } { 2 } \iint _ { A } \sigma ( \omega ) E ^ { 2 } ( r ) \diff A = \frac { 1 } { 2 } \langle \sigma ( \omega ) \rangle E _ { A } ^ { 2 } , 
\end{align}
where $E_ { A } ^ { 2 }=\int _ { A } E ^ { 2 } ( r ) d A$ quantifies the E-field inhomogeneity across the GQD. 
The averaged absorbance under vortex illumination is then derived as
\begin{align}
\label{eqn:Loren_OVspectrum}
        \left \langle \sigma \left( \omega \right) \right \rangle
        =\frac{2q^{2}\omega}{E_{A}^2}\sum_{k_{1}k}\left | \bra{\Psi_{k_{1}}} \al+i\be \ket{\Psi_{k}} \right |^{2} I_\gamma(k_{1},k,\omega),
\end{align}
where $I_\gamma(k_{1},k,\omega)=\gamma \frac{ f\left( E_{k}\right)-f\left(E_{k_{1}}\right) }{\left (E_{k_{1}k}-\hbar \omega\right )^{2}+\gamma^{2}}$, 
$\gamma$ is the Lorentzian linewidth and $f(E)$ the zero-temperature Fermi-Dirac distribution.
The matrix element $\bra{\Psi_{k_{1}}} \al+i\be \ket{\Psi_{k}}$ describes vortex-induced transitions, with $E_{k_1k} \equiv E_{k_1}-E_k$ the transition energy.

\begin{figure}[htb]
        \centering
        \includegraphics[width=0.9\columnwidth]{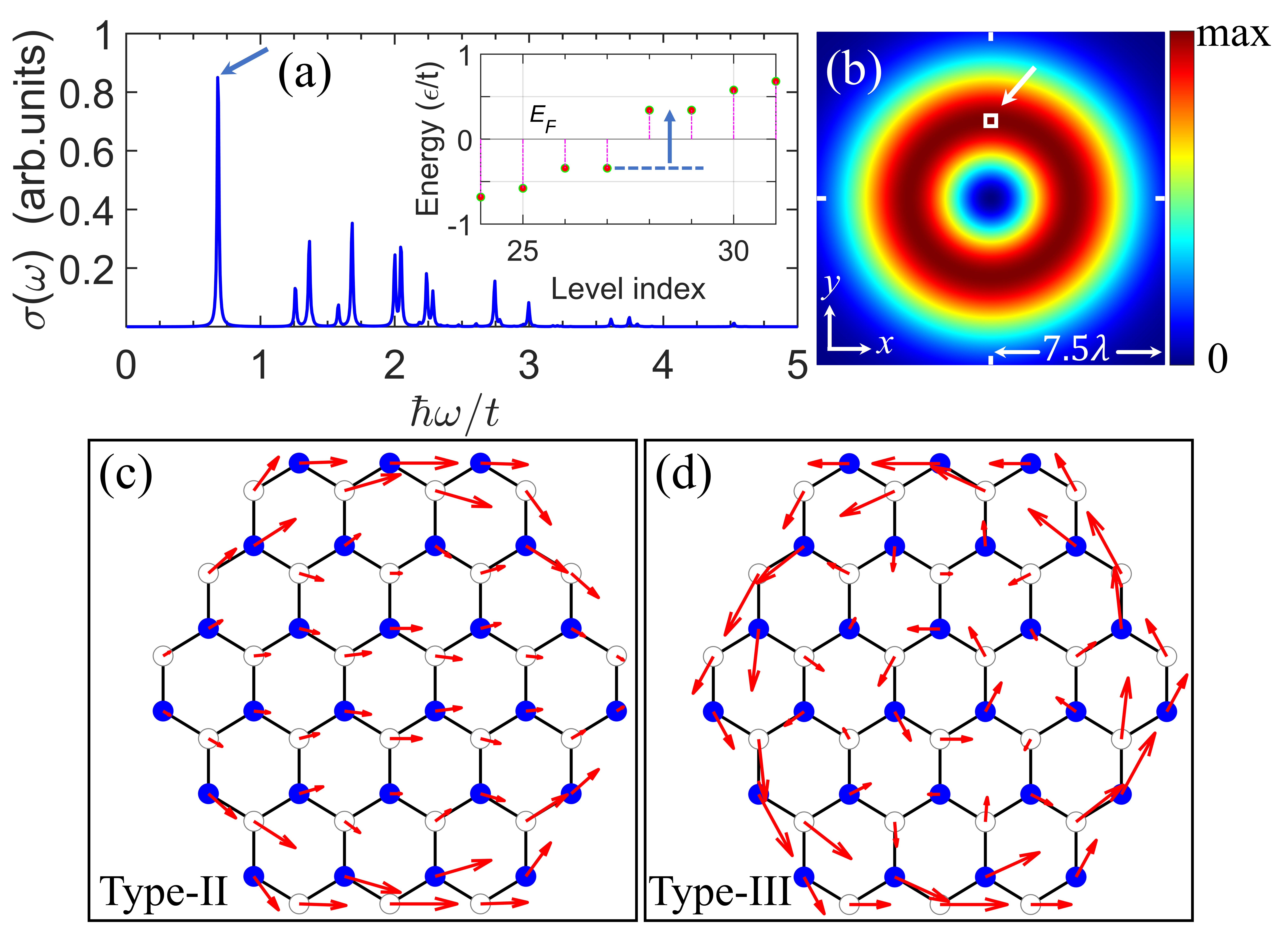}
        \caption{
        Vortex-driven responses in H-GQD ($\lambda=672$ nm). 
        (a) Absorption spectrum with zoomed energy levels (boxed position). 
        (b) $\rm LG_{01}$ beam intensity profile. 
        (c, d) Snapshots at time $t=T/4$ of (c) oscillatory type-II and (d) quasi-steady type-III current distributions. 
        Arrows in (a) and (b) mark the excitation mode and H-GQD position, respectively.
        }
        \label{fig:1}
\end{figure}

\textit{Results}---
Applying our framework, we probe vortex-driven responses in GQDs—photocurrents and angular momentum transfer. 
We select a hexagonal GQD (H-GQD, 54 atoms) for its high symmetry and experimental accessibility—a widely adopted prototype in quantum dot studies \cite{Pohle2018Symmetry}, with structure shown in Fig.~\ref{fig:1}(c, d).
We consider an H-GQD irradiated by an $\rm LG_{01}$ ($p=0,l=1$) vortex beam, positioned as in Fig.~\ref{fig:1}(b) with the characteristic length ($L_{\rm GQD}$) satisfying $L_{\rm GQD} \ll \lambda_{\rm vortex}$.
The absorption spectrum [Eq.~(\ref{eqn:Loren_OVspectrum}), Fig.~\ref{fig:1}(a)] uses a Lorentzian linewidth $\gamma = 0.02$ eV. 
We focus on the low-frequency peak (arrowed, $\hbar\omega = 0.684t = 1.847$ eV, $\lambda = 672$ nm) corresponding to lower-energy transitions. 
Its inset shows: (i) TB energy levels near the Fermi level ($E_F$), (ii) electronic transitions ($26,27 \to 28,29$), and (iii) degenerate excited states.

From the arrowed transitions in Fig.~\ref{fig:1}(a), we compute the time-dependent wavefunctions under resonant $\rm LG_{01}$ irradiation via Eq.~(\ref{eqn:Psi(t)}). 
This directly yields the photoinduced currents (type-II and type-III) through Eq.~(\ref{eqn:J_typeII_typeIII}).

Type-II currents oscillate synchronously with the in-plane E-field, while type-III currents form quasi-steady vortex around the H-GQD center—signifying electrons acquiring $z$-directional OAM that manifests as IFE. 
Their spatial distributions at $T/4$, where $T$ is the period, are shown in Fig.~\ref{fig:1}(c, d). 
The average direction of type-II currents aligns with $\bm{E}$, obeying $\bm{\bar J} = \sigma \bm{E}$. The effective scalar conductivity $\sigma$ arises from the H-GQD's hexagonal symmetry. 
Conversely, type-III currents require degenerate excited states [Eq.~(\ref{eqn:J_typeIII})], confirming that degeneracy governs OAM transfer during optical transitions. 
This constraint is consistent with SAM transfer under CPL \cite{Xu2024Photoinduced}.

\begin{figure}[htb]
        \centering
        \includegraphics[width=0.9\columnwidth]{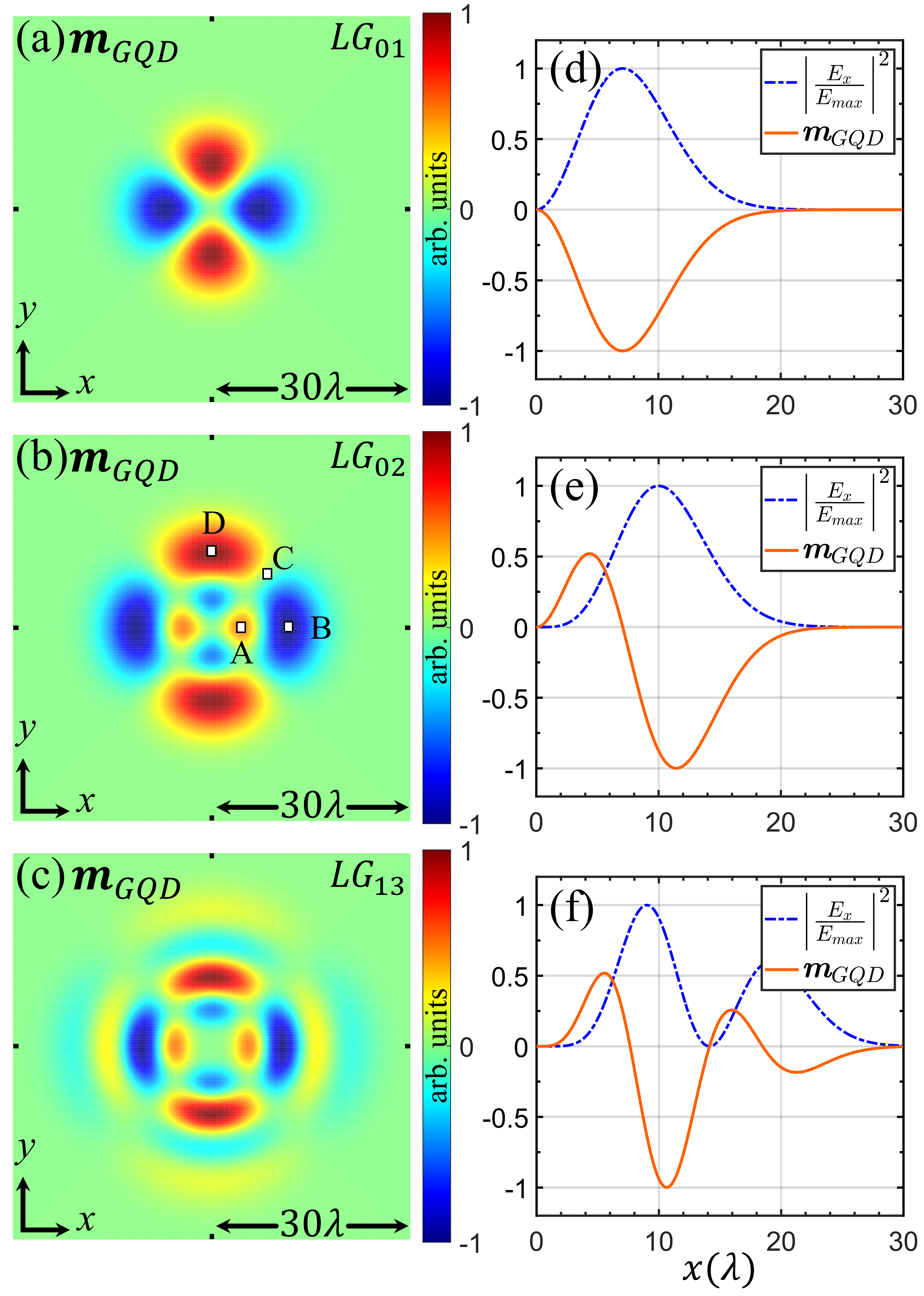}
        \caption{
        Position-resolved OMM ($\bm{m}_{\rm GQD}$) of H-GQD under ${\rm {LG}}_{pl}$ beams at fixed irradiation time. 
        (a)-(c) Distributions of $\bm{m}_{\rm GQD}$ for (a) $\rm LG_{01}$, (b) $\rm LG_{02}$, (c) $\rm LG_{13}$. 
        (d)-(f) Corresponding $x$-displacement dependencies: $\bm{m}_{\rm GQD}$ (solid) and E-field energy density (dashed). 
        Labels A-D in (b) mark positions analyzed in Fig.~\ref{fig:3}.
        }
        \label{fig:2}
\end{figure}

The spatially inhomogeneous amplitude and phase of the vortex beam induce strong position dependence on the generation of OAM, i.e. IFE.
From Eq.~(\ref{eqn:J_typeIII}), the generated orbital magnetic moment (OMM) derived as
$ \bm { m }  _ {\rm {GQD} } = \frac {1 } { 2 }  \sum _ { n } \bm{R} _ { n } \times \bm{J}_{n}^{\rm{type\mbox{-}\!III}}$ \cite{Xu2024Photoinduced}.
To quantify this position dependence, we systematically displace the H-GQD within the beam field and compute OMMs via $\bm{m}_{\rm {GQD}}$ at fixed irradiation time (Fig.~\ref{fig:2}).
Results for $\rm LG_{01}$, $\rm LG_{02}$, and $\rm LG_{13}$ modes are shown in Fig.~\ref{fig:2}(a)-(c). 
Crucially, the absorption spectrum remains invariant under GQD's shifts due to the subwavelength dimension $L_{\rm {GQD}} \ll \lambda$, consistently satisfying current-generation criteria. 
Singularity-related spectral modifications are briefly discussed in SM \cite{Supplementary}.

The OMM ($\bm{m}_{\rm GQD}$) distributions show radial/azimuthal sign reversals, with radial oscillation periods matching the LG nodal index $p$.
$\bm{m}_{\rm GQD}$ along the $x$-displacement path and the concomitant E-field energy variation are plotted in parallel panels Fig.~\ref{fig:2}(d)-(f) for each beam mode.
The sign reversal of $\bm{m}_{\rm GQD}$ occurs within the peak-width regions of E-field energy for all modes except $\rm LG_{01}$ and central ${\rm LG}_{p1}$. 
For $\rm LG_{01}$, no radial sign reversal is observed due to extreme beam confinement near the phase singularity. 
This absence of reversal similarly characterizes the innermost regions of ${\rm LG}_{p1}$ modes.

This anomalous IFE is not limited to model LG vortex beams but extends to other optical vortices \cite{Supplementary}.

For circularly polarized vortex beams, the spatial profile of the OMM mirrors the E-field energy distribution, exhibiting monopolar characteristics. 
This stems from optical SAM dominance, which masks light's OAM-induced magnetic moment features during absorption, as analytically derived in Appendix B.

The observed periodic magnetic moment reversal deviates fundamentally from theoretical predictions.
This reveals a critical paradox: the GQD acquires negative OAM when irradiated by vortex beams carrying positive OAM. 
We resolve this through our mechanism for negative OAM transfer, confirming the self-consistency of our findings.

\begin{figure}[htb]
        \centering
        \includegraphics[width=1\columnwidth]{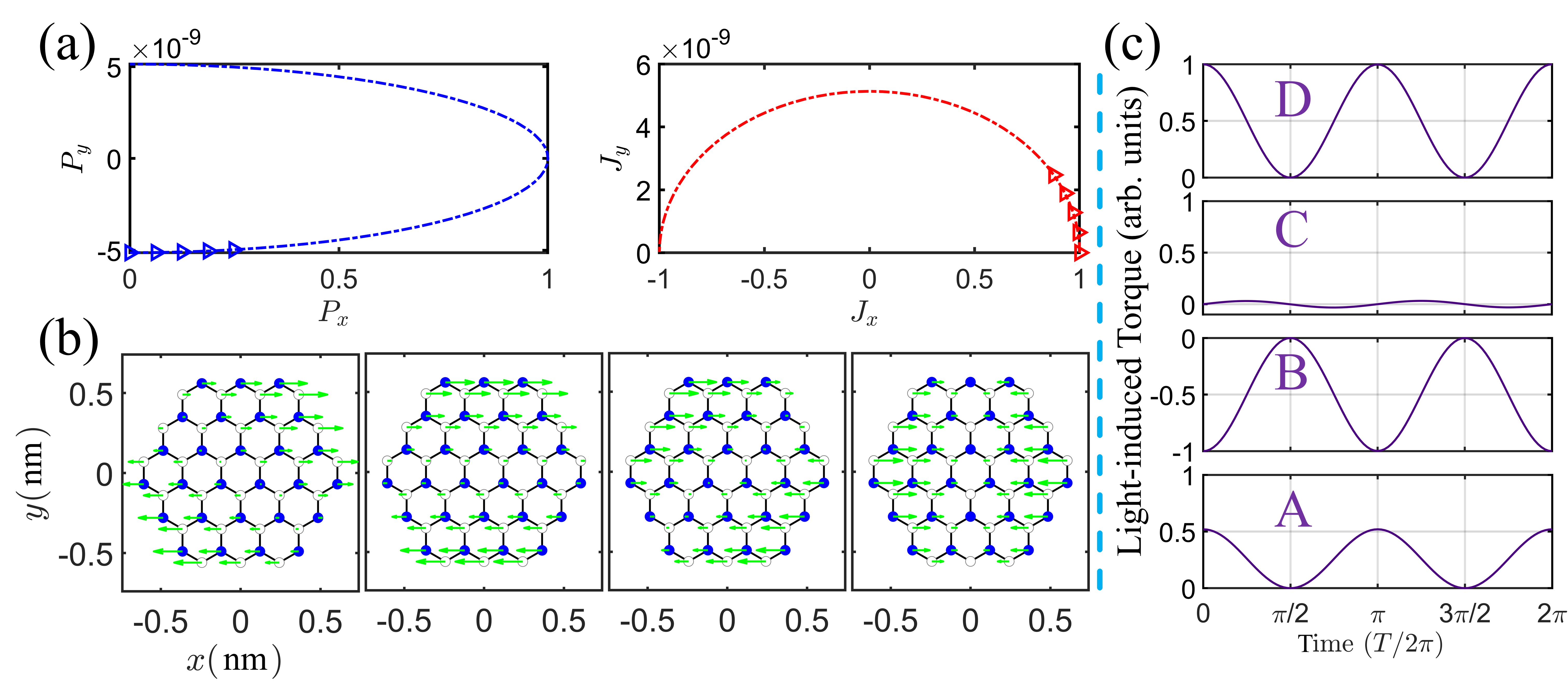}
        \caption{
        (a) Trajectories of charge polarization vector $\bm{P}$ and type-II current $\bm{J}$ over $0$--$T/2$ under $\rm LG_{02}$ irradiation at position A [Fig.~\ref{fig:2}(b)]. 
        (b) Relative variant E-field $\Delta\bm{E}(\bm R,t) = \bm E (\bm R,t) - \bm E ( \bm {R}_c ,t)$ with respect to H-GQD center $\bm {R}_c$, at snapshots: $t = T/8, T/4, 3T/8, T/2$ (left-right). 
        (c) Light-induced torque on electrons over one period ($\rm LG_{02}$), at positions A-D [Fig.~\ref{fig:2}(b)].
        }
        \label{fig:3}
\end{figure}

\textit{Phase difference mechanism}---
First, we compute the total net-charge polarization $\bm{P}$ and type-II photoinduced current $\bm{J}$, obtained by spatial summation over the GQD during half-period evolution. 
Their vector endpoints, plotted at discrete times in Fig.~\ref{fig:3}(a), reveal dominant alignment with the $x$-polarized vortex E-field. 
Minute $y$-components persist (order $10^{-9}$), implying elliptical trajectories with extreme oblateness rather than pure linear oscillation.

The origin of these minor vertical components under linearly polarized vortex illumination lies in the beam's azimuthal phase structure. 
The vortex field's $\epn^{il\theta}$ phase factor induces position-dependent phase shifts between points at different polar angles $\theta$. 
This results in spatially varying E-field amplitudes at any instant. 
Figure~\ref{fig:3}(b) quantifies the field variance $\Delta\bm{E}$ relative to the H-GQD center's field $\bm E(\bm {R}_c ,t)$ at four temporal snapshots within half a period, revealing rotational dynamics: 
the $\Delta\bm{E}$ maxima undergo anticlockwise rotation—matching type-III current vortices.

Inhomogeneous charge polarization provides a quantitative interpretation bridge. 
In electromagnetic theory, the polarized charge density $\rho^{\rm var}(\bm{R},t)$ relates to E-field spatial variation through $\rho^{\rm var}(\bm{R},t)= \bm {\nabla} \cdot \bm{P}(\bm{R},t) = \chi \bm {\nabla} \cdot \bm{E}(\bm R,t)$, where $\chi$ denotes polarizability. 
For our mesoscopic GQD system, we express this at discrete atomic sites as $\rho^{\rm var}(\bm{R},t)= i\chi \Delta\bm{E}(\bm R,t)$. 
Due to the finite GQD size, the E-field also induces boundary polarization charge  $\rho^{\rm B}(\bm{R},t) = \chi\bm{E}(\bm{R}_c,t)$.
The bulk ($\rho^{\rm var}$) and boundary ($\rho^{\rm B}$) charges yield the total polarization vector $\bm{P}$ shown in Fig.~\ref{fig:3}(a), 
respectively generating its circular (rotational) and linear ($x$-directional) components.
Since $\partial_t\bm{P}$ equals the polarized current, the rotational component corresponds to type-III current, while the linear component gives the linear oscillatory current; together they constitute type-II current.

Both $\bm{P}$ and $\bm{J}$ exhibit anticlockwise rotation—mirroring $\Delta\bm{E}$—confirming the phase-difference-driven rotation mechanism under linearly polarized vortex illumination.

The light-induced torque $\bm{M} =\bm{P}\times\bm{E}$ shows $2\omega$-sinusoidal behavior [Fig.~\ref{fig:3}(c)] due to the $x$-polarized $\bm{E}$-field and elliptical $\bm{P}$-trajectory. 
At different positions A-D [Fig.~\ref{fig:2}(b)], the rotation sense of $\Delta\bm{E}$, $\bm{P}$, and $\bm{M}$ determines magnetic moment sign alternation between positive/negative regimes.

\textit{Eigenmode decomposition analysis}---
Phenomenological explanation of negative angular momentum acquisition leaves a core paradox: 
why does a GQD in an ${\rm LG}_{pl}$ mode (each photon carrying $l\hbar$ OAM) acquire reversed angular momentum in specific regions—seemingly violating angular momentum conservation?
This anomaly stems from off-axis positioning: the radiation field experienced by the GQD constitutes a superposition of ${\rm LG}_{pl}$ eigenmodes centered at its location. 
This mode mixing drives anomalous angular momentum transfer, while eigenmode decomposition as a methodology is reported in Refs \cite{Quinteiro2010Electronic, Barnett2022Optical}.

\begin{figure}[htb]
        \centering
        \includegraphics[width=1\columnwidth]{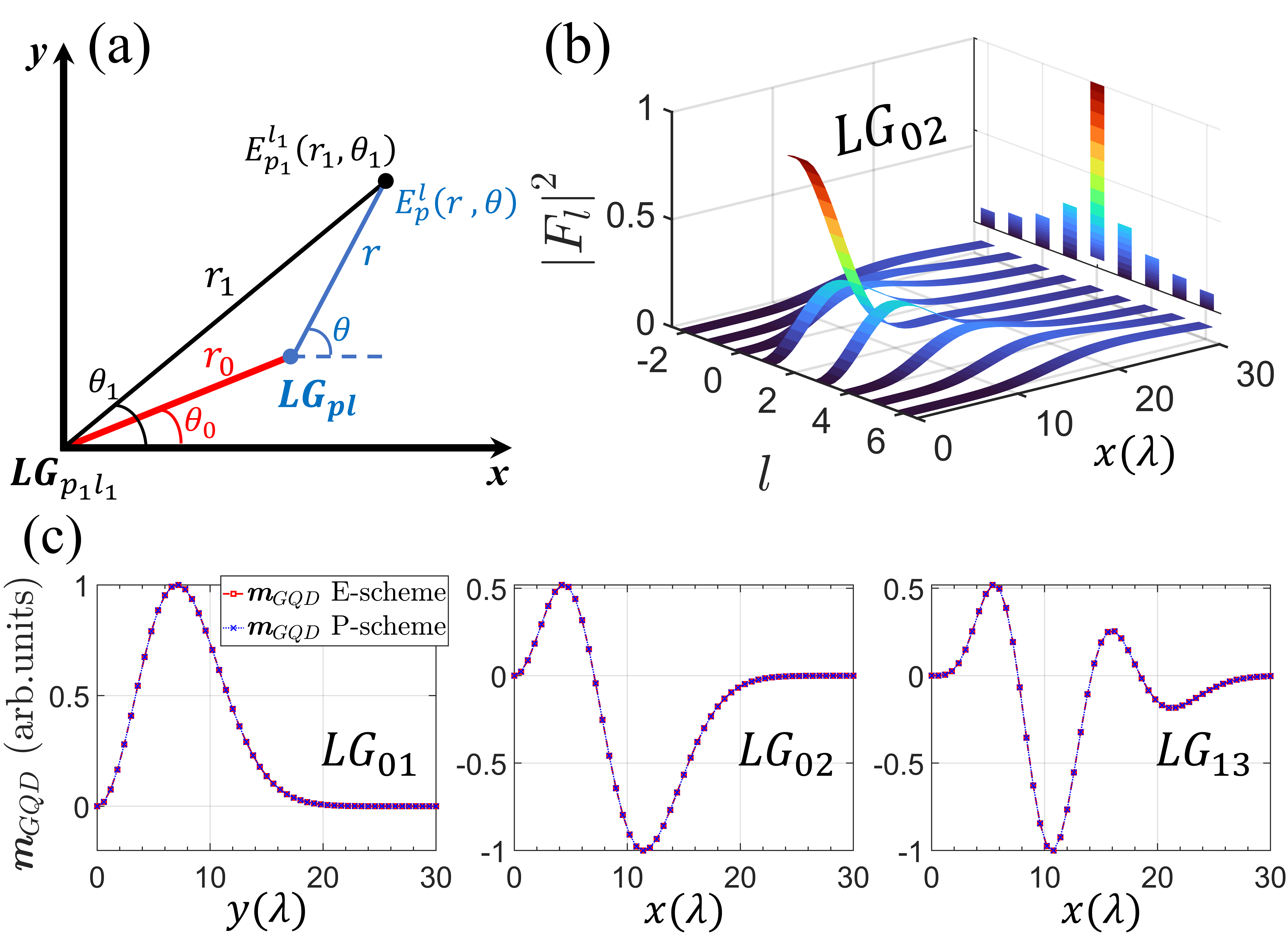}
        \caption{
        (a) Coordinate transformation schematic: origin-centered ${\rm LG}_{p_1l_1}$ field expressed as off-axis ${\rm LG}_{pl}$ superposition [Eq.~(\ref{eqn:decom})]. 
        (b) Decomposition weights $|F_l|^2$ for $x$-displaced ${\rm LG}_{pl}$ modes derived from origin-centered $\rm LG_{02}$, plotted versus displacement $x$ and OAM index $l$. 
        (c) Position-resolved absorbed magnetic moment $\bm{m}_{\rm GQD}$ in H-GQD under $\rm LG_{01}$, $\rm LG_{02}$, $\rm LG_{13}$ modes, computed via E-scheme and P-scheme.
        }
        \label{fig:4}
\end{figure}

All ${\rm LG}_{pl}$ modes sharing a common axis constitute a complete orthogonal basis [$p\in(0,\infty)$, $l\in(-\infty,\infty)$] \cite{Riley2006Mathematical}. 
Figure~\ref{fig:4}(a) illustrates how the E-field $E_{p_1}^{l_1}(r_1,\theta_1)$ of an origin-centered ${\rm LG}_{p_1l_1}$ mode decomposes into a superposition of off-axis ${\rm LG}_{pl}$ fields $E_p^l(r,\theta)$, centered at $\bm{R}_0=(r_0,\theta_0)$. 
This coordinate transformation is expressed as
\begin{align}
\label{eqn:decom}
        E _ { p _ { 1 } } ^ { l _ { 1 } } ( r _ { 1 } , \theta _ { 1 } ) = \sum _ { p l } F _ { p l } \, E _ { p } ^ { l } ( r , \theta ), 
\end{align}
where $F_{pl}=\langle E_{ p }^{ l }( r , \theta ) \,|\, E_{ p_{ 1 }}^{ l_{ 1 }} ( r_{ 1 }, \theta_{ 1 }) \rangle$ are the expansion coefficients, and the geometric relationships satisfy:
$r_{ 1 }^{ 2 }=r_{ x }^ { 2 }+r_{ y }^ { 2 }$, $\tan \theta_{ 1 } =\frac {r_{ y }}{r_{ x }}$, where $r_{ x }=r \cos \theta + r _ { 0 } \cos \theta _ { 0 }$, $r_{ y }=r \sin \theta + r _ { 0 } \sin \theta _ { 0 }$.

Using Eq.~(\ref{eqn:decom}), we compute the off-axis decomposition weights for an origin-centered $\rm LG_{02}$ beam. 
Figure~\ref{fig:4}(b) shows the $l$-resolved distribution of $|F_l|^2$ (where $|F_l|^2 = \sum_p |F_{pl}|^2$) for ${\rm LG}_{pl}$ modes displaced along $+x$. 
Crucially, the $|F_l|^2$ versus $x$ curves exhibit mirror symmetry about $l=2$—a signature of the parent $\rm LG_{02}$ mode's topological charge conservation.

Within this multi-mode superposition framework, we derive the expectation of absorbed angular momentum for off-center GQDs. 
First, analyzing GQD electron energy absorption via Eqs.~(\ref{eqn:joule})-(\ref{eqn:Loren_OVspectrum}), the total absorbed photon energy at frequency $\omega$ is given by
\begin{align}
\label{eqn:ab_energy_A}
        \mcal{E}_{\rm LP}\propto \omega \sum_{k_{1}k}\left | \bra{\Psi_{k_{1}}} x E_{p_1}^{l_1}(r_1,\theta_1) \ket{\Psi_{k}} \right |^{2} I_\gamma(k_{1},k,\omega).
\end{align}
Rewriting Eq.~(\ref{eqn:ab_energy_A}) via Eq.~(\ref{eqn:decom}) yields
\begin{align}
\label{eqn:ab_energy_B}
        \mcal{E}_{\rm LP}\propto \omega \sum_{p_{a}l_{a}} \sum_{p_{b}l_{b}} F_{p_{a}l_{a}}^{*} F_{p_{b}l_{b}} \mcal{ E }_{\rm LP}^{p_{a}l_{a}, \, p_{b}l_{b}}, 
\end{align}
where $\mcal{ E }_{\rm LP}^{p_{a}l_{a}, \, p_{b}l_{b}}$ denotes the coupled-mode absorbed energy, given by 
$$\mcal{ E }_{\rm LP}^{p_{a}l_{a}, \, p_{b}l_{b}}=\sum_{k_{1}k} [x E_{p_a}^{l_a}(r,\theta) ]_{k_{1}k}^{*} [ x E_{p_b}^{l_b}(r,\theta) ]_{k_{1}k}  I_\gamma(k_{1},k,\omega).$$
This represents a decomposition of the absorption process. 
Equation~(\ref{eqn:ab_energy_B}) reveals that the total absorbed energy $\mathcal{E}_{\rm LP}$ originates not only from intramode absorption ($p_a=p_b, l_a=l_b$), but also from intermode couplings.   

For ${\rm LG}_{pl}$ eigenmodes in GQD regions, the angular-momentum-to-energy flux ratio is $\frac {L_{z}} {W}=\frac{l} {\omega}$ \cite{Allen1992Orbital}.
And the corresponding ratio for intermode couplings is derived rigorously as  
\begin{align}
\label{eqn:coup_ratio}
        \frac{(L_z)_{\rm couple}}{W_{\rm couple}} = \frac{l_a + l_b}{2\omega},
\end{align}
with full derivation in Appendix A. 
When $l_a=l_b$, this coupled-mode ratio reduces to the single-mode form ($l/\omega$).
Consequently, the expected absorbed angular momentum is
\begin{align}
\label{eqn:expected_L}
        \langle \bm{L} \rangle _{\rm LP} \propto \sum_{p_{a}l_{a}} \sum_{p_{b}l_{b}} F_{p_{a}l_{a}}^{*} F_{p_{b}l_{b}} \mcal{ E }_{\rm LP}^{p_{a}l_{a}, \, p_{b}l_{b}} (l_a + l_b).
\end{align}

The OAM absorption via type-III currents [Figs.~\ref{fig:1}(d) and \ref{fig:2}] must equal the expectation value of beam-transferred angular momentum. 
Using Eq.~(\ref{eqn:expected_L}), the expected absorbed magnetic moment is computed for H-GQD displaced along $+y$ ($\rm LG_{01}$) and $+x$ ($\rm LG_{02}$, $\rm LG_{13}$) against type-III current calculations in Fig.~\ref{fig:4}(c).
Hereafter, eigenmode-absorption (E-scheme) and photoinduced-current (P-scheme) approaches show excellent agreement across irradiation conditions. 
This establishes that the bipolar magnetic moment distribution in Fig.~\ref{fig:2} stems from off-axis beam superposition effects.

This eigenmode decomposition method also provides a novel approach to determining the absorbed angular momentum in quantum-confined systems coupled with optical vortices.

\textit{Conclusions}---
We establish a framework for vortex-driven photoresponses in GQDs, extending our CPL approach \cite{Xu2024Photoinduced} to arbitrary vortex configurations via TB model and time-dependent perturbation theory.
Crucially, degenerate-state-selective type-III currents generate quasi-steady rotational patterns that manifest an anomalous IFE with emergent magnetic moments.

Spatially resolved measurements reveal periodic polarity reversal of photoinduced magnetic moments under linearly polarized vortex illumination—exhibiting fundamental deviation from the direction of incident OAM. 
This bipolar distribution, rigorously validated through time-resolved charge polarization, stems from phase-difference mechanism of vortex E-field.
The apparent violation of angular momentum conservation in this anomalous IFE is also self-consistently explained by the coaxial eigenmode superpositions via off-axis field decomposition.
These findings reveal a counterintuitive angular momentum conversion paradigm in quantum-confined systems, advancing fundamental understanding of vortex-nanostructure interactions.

\textit{Acknowledgments}---
This work was partially supported by the National Natural Science Foundation of China under Grant No. 12347101.

\bibliography{Vortex_IFE_GQD_Ref}
\bibliographystyle{apsrev4-2}

\newpage
\appendix

\onecolumngrid
\section*{End Matter}
\twocolumngrid

\setcounter{equation}{0}
\renewcommand{\theequation}{A\arabic{equation}}

\emph{Appendix A: Angular-momentum-to-energy-flux ratio in superposed LG beams}--- 
This derivation gives the corresponding ratio for coherently superposed ${\rm LG}_{pl}$ beams in Eq.~(\ref{eqn:coup_ratio}). 
Following an approach analogous to Allen \cite{Allen1992Orbital}, two coaxial, $x$-polarized LG beams can be expressed via the vector potential as
\begin{align}
\label{eqn:v_potential}
        \bm {A} = \hat { \bm{ x } } u ( r , \theta , z ) \exp [ i ( k z - \omega t ) ],
\end{align}
where $\hat{\bm{x}}$ denotes the unit vector along the $x$-axis. 
The total complex field amplitude $u (r, \theta, z)=u_a+u_b$ comprises two LG modes: $u_{a} = E_{p_{a}}^{l_{a}}(r, z) \exp (il_{a}\theta)$ and $u_{b} = E_{p_{b}}^{l_{b}}(r, z) \exp (il_{b}\theta)$.

Under paraxial approximation, the electromagnetic field is expressed as
\begin{align}
        \bm { E } &= \left( i \omega u \hat { \bm { x } } - c \frac { \partial u } { \partial x } \hat { \bm { z } } \right) \exp [ i ( k z - \omega t ) ] , \nt 
\label{eqn:EM_field}
        \bm { B } &= \left( i k u \hat { \bm { y } } - \frac { \partial u } { \partial y } \hat { \bm { z } } \right) \exp [ i ( k z - \omega t ) ] , 
\end{align}
where $c$ denotes the speed of light. 
Within this formalism, the time-averaged linear momentum density $\langle \bm{\mcal{P}} \rangle$ in vacuum, derived from $\real (\varepsilon_{0} \bm E \times \bm{B})$, is given by
\begin{align}
        \langle \bm{\mcal{P}} \rangle &=\frac{\varepsilon_{0}}{4} (\bm{E}^{*} \times \bm B +\bm E \times \bm{B}^{*}) \nt
\label{eqn:time_momentum}
        &= \frac{\varepsilon_{0} \omega}{4} \Big[i ( u \bm{\nabla}\!_{\perp} u^{*} - u^{*} \bm{\nabla}\!_{\perp} u) + 2k |u|^{2} \hat{\bm{z}}\Big], 
\end{align}
where $\varepsilon_{0}$ is the vacuum permittivity. Nonlinear effects emerge in this formulation. 
Substituting $u=u_a+u_b$ into Eq.~(\ref{eqn:time_momentum}) yields the azimuthal component of $\langle \bm{\mcal{P}} \rangle$ expressed as
\begin{align}
\label{eqn:P_theta}
        \langle \mcal{P}_{\hat{\bm{\theta}}} \rangle=\frac{\varepsilon_{0}\omega}{2} \Big [ \frac {l_{a}}{r} |u_{a}|^{2} + \frac{l_{b}}{r} |u_{b}|^{2} + \frac{l_{a}+l_{b}}{2r} \left(u_{a} u_{b}^{*} + u_{b} u_{a}^{*} \right) \Big ].
\end{align}
Therefore, the time-averaged angular momentum density component along the propagation axis ($z$), derived from $\bm{\mcal{L}} = \bm{R} \times \bm{\mcal{P}}$, is given by
\begin{align}
        \langle  \mcal{L}_{\hat{\bm{z}}} \rangle = r \langle \mcal{P}_{\hat{\bm{\theta}}} \rangle = & \frac{\varepsilon_{0}\omega}{2} \Big [ l_{a} |u_{a}|^{2} + l_{b} |u_{b}|^{2} \nt
\label{eqn:time_Lz}
        & + \frac{l_{a}+l_{b}}{2} \left(u_{a} u_{b}^{*} + u_{b} u_{a}^{*} \right) \Big ].
\end{align}
Additionally, the time-averaged energy density ($\langle w \rangle$) of this superposed beam is defined as
\begin{align}
        \langle w \rangle &=\frac{\varepsilon_{0}}{4} (\bm{E} \cdot \bm{E}^{*} + c^{2} \bm{B} \cdot \bm{B}^{*} ) \nt
\label{eqn:time_energy}
        &=\frac{\varepsilon_{0} \omega^{2} }{2} \Big [ |u_{a}|^{2} + |u_{b}|^{2} + \left(u_{a} u_{b}^{*} + u_{b} u_{a}^{*} \right) \Big ].
\end{align}

The ratio of angular momentum flux to energy flux along the propagation direction is then obtained by comparing the corresponding terms in Eqs.~(\ref{eqn:time_Lz}) and (\ref{eqn:time_energy}), 
yielding the explicit form given in Eq.~(\ref{eqn:coup_ratio}).

\setcounter{equation}{0}
\renewcommand{\theequation}{B\arabic{equation}}

\vspace{0.5cm}
\emph{Appendix B: GQDs driven by circularly polarized optical vortices}--- 
Circularly polarized vortex beams carry coexisting SAM and OAM, generating distinct magnetization responses in GQDs versus linear polarization.  
The E-field of a vortex beam with circular polarization ($\sigma\!_s = \pm 1$) is
\begin{align}
\label{eqn:E(Rt)_CP}
        \bm {E}_{\rm CP}^{\sigma\!_s}  ( \bm  R  , t ) =  E ( r , z ) \exp ( i l \theta ) \exp [ i ( k z - \om t ) ] \hat {\bm {\epn}}_{\sigma\!_s} ,
\end{align}
where the Jones vector $\hat {\bm {\epn}}_{\sigma\!_s}=\hat{\bm{x}} + i\sigma\!_s \hat{\bm{y}}$ governs SAM transfer, with the normalization factor $1/\sqrt{2}$ omitted for simplicity.

Analogous to the derivation of Eq.~(\ref{eqn:Psi(t)}), the time-dependent electronic wavefunction of the GQD in the optical field is expressed as
\begin{align}
        \Psi_{\rm CP}^{\sigma\!_s} ( t ) &= \Psi _ { k } \epn^ { - i \omega _ { k } t } + \sum _ { k _ { 1 } } S _ { k _ { 1 } } ( t ) \Big [ (x + i \sigma\!_s y) E ( r ) \epn ^ { i l \theta } \Big ] _ { k _ { 1 } k } \nt
        &\;\;\;\; \label{eqn:Psi(t)_CP}
        \times \Psi _ { k _ { 1 } } \epn ^ { - i \omega _ { k _ { 1 } } t } .
\end{align}
The operator $(x + i \sigma\!_s y) E(r) \epn^{i l \theta}$ yields the form $\al_c+i\be_c$, where
\begin{align}
        \al_c =E(r) (x\cos l\theta - \sigma\!_s y \sin l\theta), \nt
\label{eqn:al_be_CP}
        \be_c =E(r) (x\sin l\theta + \sigma\!_s y \cos l\theta).
\end{align}
Replacing $\alpha \to \alpha_c$ and $\beta \to \beta_c$ in Eqs.~(\ref{eqn:J_typeII_typeIII}) and (\ref{eqn:Loren_OVspectrum}) yields the type-II/III currents and absorption spectrum under circularly polarized vortex beams. 
This substitution extends consistently to the magnetic moment calculation ($\bm{m}_{\rm GQD}$).

Under identical configurations, type-II/III currents under circular polarization retain Results-section characteristics. 
However, the magnetic moment distribution fundamentally differs: it replicates the E-field energy profile without periodic polarity reversal, as shown by blue dashed curves in Fig.~\ref{fig:2}(d)-(f).

Distinct from linear polarization, SAM-induced magnetic moments dominate OAM contributions by orders of magnitude in GQDs. 
Figure~\ref{fig:3}(a) reveals the mechanism: circularly polarized vortex field generates comparable vertical/horizontal components via field rotation—unlike linear polarization's weak $P_y \sim 10^{-9}$ from $\epn^{il\theta}$ phase. 
Consequently, optical spin torque ($\bm{M} =\bm{P}\times\bm{E}$) masks persistent but weaker orbital torque. 
The SAM-dominant magnetic moment distribution thus replicates the E-field energy profile. 

Extending the eigenmode decomposition framework, we derive the expectation value of absorbed angular momentum under circularly polarized LG beams.
The total absorbed photon energy at frequency $\omega$ is given by
\begin{align}
\label{eqn:ab_energy_CP}
        \mcal{E}_{\rm CP} \propto \omega \sum_{p_{a}l_{a}} \sum_{p_{b}l_{b}} F_{p_{a}l_{a}}^{*} F_{p_{b}l_{b}} \mcal{ E }_{\rm CP}^{p_{a}l_{a}, \, p_{b}l_{b}}, 
\end{align}
where $\mcal{ E }_{\rm CP}^{p_{a}l_{a}, \, p_{b}l_{b}}$ is the coupled-mode absorbed energy, given by 
\begin{widetext}
$$\mcal{ E }_{\rm CP}^{p_{a}l_{a}, \, p_{b}l_{b}}=\sum_{k_{1}k} \Big [(x + i \sigma\!_s y) E_{p_a}^{l_a}(r,\theta) \Big]_{k_{1}k}^{*}  \Big [(x + i \sigma\!_s y) E_{p_b}^{l_b}(r,\theta) \Big]_{k_{1}k}  I_\gamma(k_{1},k,\omega).$$
\end{widetext}

We now compute the ratio of angular momentum to energy flux for circular polarization within the GQD region. 
For two coaxial, coherently superposed circularly polarized LG beams, the time-averaged linear momentum density follows the Appendix A methodology, yielding  
\begin{align}
        \langle \bm{\mcal{P}} \rangle _{\rm CP} =& \frac{\varepsilon_{0} \omega}{4} \Big[i ( u \bm{\nabla}\!_{\perp} u^{*} - u^{*} \bm{\nabla}\!_{\perp} u) + 2k |u|^{2} \hat{\bm{z}} \nt
\label{eqn:time_momentum_CP}
        & + \sigma\!_s \left( \frac { \partial } { \partial y } \hat{\bm{x}} - \frac { \partial } { \partial x } \hat{\bm{y}} \right) |u|^{2} \Big]. 
\end{align}
The operator within the third right-hand-side term rewrites under coordinate system transformation as 
\begin{align}
\label{eqn:co_trans}
        \frac { \partial } { \partial y } \hat{\bm{x}} - \frac { \partial } { \partial x } \hat{\bm{y}} = \frac{1}{r} \frac { \partial } { \partial \theta } \hat{\bm{r}} - \frac { \partial } { \partial r } \hat{\bm{\theta}}.
\end{align}
By applying the decomposition $u=u_a+u_b$, the azimuthal component of $\langle \bm{\mcal{P}} \rangle _{\rm CP}$ is derived as
\begin{align}
        \langle \mcal{P}_{\hat{\bm{\theta}}} \rangle _{\rm CP}=& \frac{\varepsilon_{0}\omega}{2} \Big [ \frac {l_{a}}{r} |u_{a}|^{2} + \frac{l_{b}}{r} |u_{b}|^{2} \nt
\label{eqn:P_theta_CP}
        & + \frac{l_{a}+l_{b}}{2r} \left(u_{a} u_{b}^{*} + u_{b} u_{a}^{*} \right) - \frac{\sigma\!_s}{2} \frac { \partial |u|^{2}} { \partial r } \Big ].
\end{align}
Therefore, the time-averaged angular momentum density along the propagation ($z$) axis is
\begin{align}
        \hspace{-0.3em}
        \langle \mcal{L}_{\hat{\bm{z}}} \rangle _{\rm CP}\!= & \frac{\varepsilon_{0}\omega}{2} \Big [ \left( l_{a} - \frac{\sigma\!_s r}{2} \frac {\partial} { \partial r } \right) \! |u_{a}|^{2} 
        +  \left( l_{b} - \frac{\sigma\!_s r}{2} \frac {\partial} { \partial r } \right) \! |u_{b}|^{2} \nt
\label{eqn:time_Lz_CP}
        & + \left(\frac{l_{a}+l_{b}}{2} - \frac{\sigma\!_s r}{2} \frac {\partial} { \partial r } \right) \! \left(u_{a} u_{b}^{*} + u_{b} u_{a}^{*} \right) \Big ].
\end{align}
Additionally, the time-averaged energy density for superposed circularly polarized beam retains the form of Eq.~(\ref{eqn:time_energy}).
Within GQDs, comparison of Eq.~(\ref{eqn:time_Lz_CP}) with Eq.~(\ref{eqn:time_energy}) reveals that 
the angular-momentum-to-energy flux ratio lacks a simple scalar proportionality—in contrast to linear polarization [Eq.~(\ref{eqn:time_Lz})]—owing to spin-induced gradient terms \cite{Allen2000ThePoynting}.

Nevertheless, the absorbed coupled angular momentum in GQDs maintains analytical tractability. 
To enable theoretical derivation, we introduce the energy absorption density factor $\tau_{u_a,u_b}$ governing this radiative process. 
The coupled absorption energy is then expressed as $\mcal{ E }_{\rm CP}^{p_{a}l_{a}, \, p_{b}l_{b}} = \iint _ { A } \tau_{u_a,u_b} \diff A$, where $\tau_{u_a,u_b}$ vanishes outside the GQD area $A$. 
This formulation facilitates expression of the net absorbed coupled angular momentum under beam superposition as
\begin{widetext}
\begin{align}
        \bm {L}_{\rm CP}^{p_{a}l_{a}, \, p_{b}l_{b}} =& \iint _ { A } \tau_{u_a,u_b} \frac{ \langle \mcal{L}_{\hat{\bm{z}}} \rangle _{\rm CP}^{\rm {couple}}}{\langle w \rangle _{\rm {couple}}} \diff A 
        =\iint _ { A } \tau_{u_a,u_b} \frac{ \left(\frac{l_{a}+l_{b}}{2} - \frac{\sigma\!_s r}{2} \frac {\partial} { \partial r } \right) \left(u_{a} u_{b}^{*} + u_{b} u_{a}^{*} \right) }{\omega \left(u_{a} u_{b}^{*} + u_{b} u_{a}^{*} \right)} \diff A \nt
        =&\mcal{ E }_{\rm CP}^{p_{a}l_{a}, \, p_{b}l_{b}} \frac{l_{a}+l_{b}}{2 \omega} -\frac{\sigma\!_s}{2 \omega} \iint _ { A } \frac{ \tau_{u_a,u_b} r^2 }{\left(u_{a} u_{b}^{*} + u_{b} u_{a}^{*} \right)} \frac {\partial} { \partial r } \left(u_{a} u_{b}^{*} + u_{b} u_{a}^{*} \right) \diff r \diff \theta \nt
\label{eqn:expected_LCP_ab}
        =&\mcal{ E }_{\rm CP}^{p_{a}l_{a}, \, p_{b}l_{b}} \frac{ l_{a}+l_{b} + 2 \sigma\!_s}{2 \omega}.
\end{align}
\end{widetext}
This derivation utilizes the approximate linear dependence: $\tau_{u_a,u_b} \propto \real(u_{a} u_{b}^{*})$, derived from Joule heating formalism.

Contrasting the derivation process of Eq.~(\ref{eqn:expected_L}), the expectation value of absorbed electronic angular momentum through Eqs.~(\ref{eqn:ab_energy_CP}) and (\ref{eqn:expected_LCP_ab}) yields 
\begin{align}
        \hspace{-0.9em} 
        \langle \bm{L} \rangle _{\rm CP} & \propto \omega \sum_{p_{a}l_{a}} \sum_{p_{b}l_{b}} F_{p_{a}l_{a}}^{*} F_{p_{b}l_{b}} \bm {L}_{\rm CP}^{p_{a}l_{a}, \, p_{b}l_{b}} \nt
\label{eqn:expected_LCP}
        & \propto \sum_{p_{a}l_{a}} \sum_{p_{b}l_{b}} F_{p_{a}l_{a}}^{*} F_{p_{b}l_{b}} \mcal{ E }_{\rm CP}^{p_{a}l_{a}, \, p_{b}l_{b}} (l_{a}+l_{b} + 2 \sigma\!_s). 
\end{align}
Mirroring the linear polarization vortex results [Fig.~\ref{fig:4}(c)], our numerical calculations for circularly polarized vortex illumination demonstrate excellent agreement for angular momentum (magnetic moment) distributions from
(i) the photoinduced-current approach, and (ii) expectation value distributions via eigenmode decomposition using Eq.~(\ref{eqn:expected_LCP}).

\end{document}


\title{Anomalous inverse Faraday effect for graphene quantum dots in optical vortices \texorpdfstring{\\}{ -- }(Supplemental Material)}

\author{Zi-Yang Xu}
\affiliation{College of Physics, Chongqing University, Chongqing 401331, China}

\author{Wei E. I. Sha}
\affiliation{College of Information Science and Electronic Engineering, Zhejiang University, Hangzhou 310027, China}

\author{Hang Xie}
\email[Corresponding author: ]{xiehang2001@hotmail.com}
\affiliation{College of Physics, Chongqing University, Chongqing 401331, China}
\affiliation{Chongqing Key Laboratory for Strongly-Coupled Physics, Chongqing University, Chongqing 401331, China}

\maketitle

\section{Quadrupole contribution}
This section establishes that in regions of the optical vortex field sufficiently distant from the singularity (specifically, for $r\gtrsim{10}^{-2}\lam $), electric dipole interactions constitute the dominant mechanism, thereby rendering quadrupole contributions negligible. 
This comparative analysis of their respective interaction strengths is conducted through the framework of optical absorption. 
Analogous to the electric dipole absorption spectrum given by Eq.~(9), the quadrupole absorption spectrum is expressed as
\begin{align}
\label{eqn:Loren_OVqpspectrum}
        \left \langle \sigma_{qp} \left( \omega \right) \right \rangle
        =\frac{2\omega}{E_{A}^2}\sum_{k_{1}k}\left | \bra{\Psi_{k_{1}}} H_{qp} \ket{\Psi_{k}} \right |^{2} I_\gamma(k_{1},k,\omega),
\end{align}
where $$H_{qp}=-\frac{1}{2} \sum_{\mu \nu }Q_{\mu \nu } \nabla\!_{\mu} E_{\nu }.$$
$Q_{\mu \nu} = q x_\mu x_\nu$ are the components of the quadrupole moment tensor, where $x_\mu$ are the components of the internal position vector $\bf x$, and $\nabla\!_{\mu}$ are the components of the gradient operator.
By comparing the relative strengths of electric dipole and quadrupole optical absorption, ${\left \langle \sigma \left( \omega \right) \right \rangle _{\rm max}}/{\left \langle \sigma_{qp} \left( \omega \right) \right \rangle _{\rm max}}$, 
calculated at various spatial positions via Eqs.~(9) and (\ref{eqn:Loren_OVqpspectrum}), the influence of quadrupole interactions can be quantitatively characterized.

\begin{figure}[!htb]
        \centering
        \includegraphics[width=1\columnwidth]{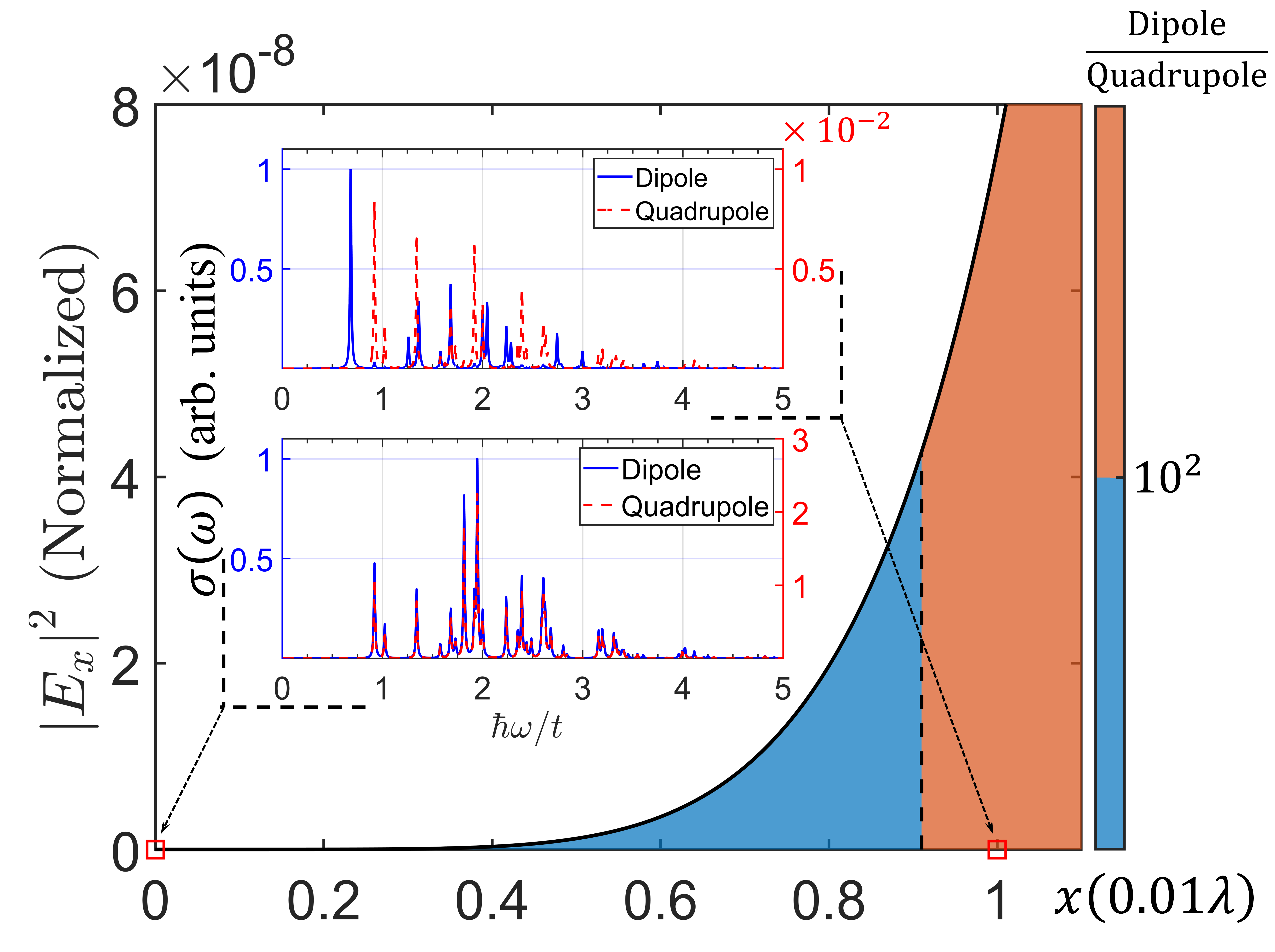}
        \caption{
        Intensity of electric dipole and quadrupole interactions between the $\rm {LG}_{13}$ beam and the H-GQD, overlaid on the radial ($x$-direction) E-field energy profile. 
        A critical boundary demarcates regions where dipole interactions are dominant (orange-shaded area). 
        Insets show the optical absorption spectra for dipole and quadrupole components at the beam singularity and at position ($0.01\lam$, 0), normalized to the local dipole absorption intensity.
        }
        \label{figsm:1}
\end{figure}

Figure~\ref{figsm:1} displays the spatial variation of the relative absorption intensities between electric dipole and quadrupole moments in the H-GQD under an $\rm LG_{13}$ beam, evaluated along the $x$-direction. 
The variation is superimposed on the radial electric field energy profile. 
A defined critical threshold (intensity ratio $10^{2}$, black dashed line) demarcates the regime where quadrupole absorption becomes negligible relative to dipole contributions, i.e., where dipole interactions dominate beyond $ 10^{-2}\lambda$ from the singularity core. 
Notably, although the region influenced by quadrupole effects expands with increasing topological charge $l$, its spatial extent remains orders of magnitude smaller than the beam's characteristic radius.

The insets in Fig.~\ref{figsm:1} compare absorption spectra for dipole and quadrupole components at the singularity and at ($0.01\lam$, 0). 
Distinct absorption behaviors are observed: at the singularity, both components show similar spectral shapes with quadrupole dominance over dipole (enhanced by increasing $l$). 
In contrast, at the dipole-dominated position ($0.01\lam$, 0), their spectral profiles diverge markedly. 
Notably, both types of spectral lines remain unchanged across the dipole-dominated (orange) region. 
In the quadrupole-active (blue) region, however, the two spectra exhibit complex variations that lie beyond the scope of this study.

\section{Magnetic moment distribution in GQDs under other optical vortices}

\begin{figure*}[htb]
        \centering
        \includegraphics[width=2\columnwidth]{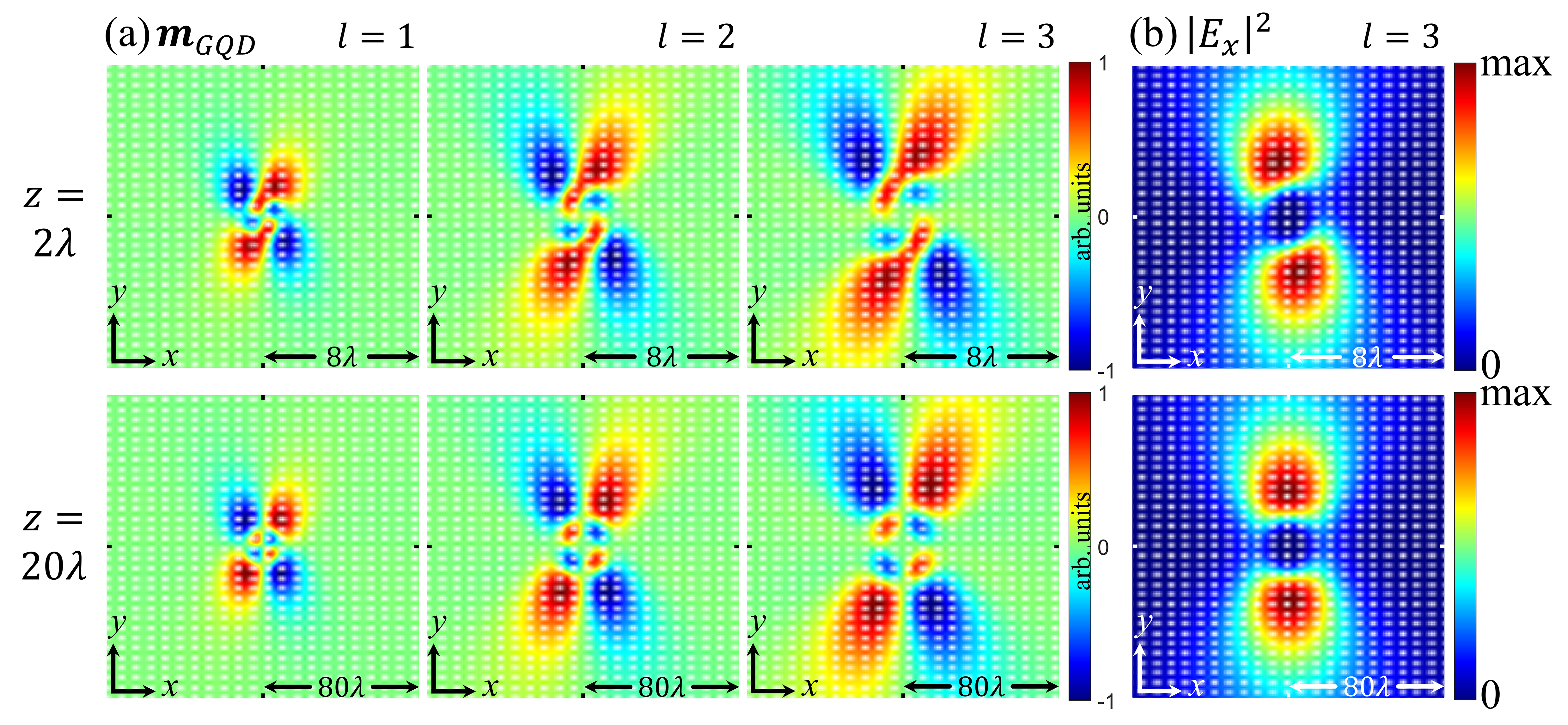}
        \caption{
        Position-resolved magnetic moment distribution ($\bm{m}_{\rm{GQD}}$) of H-GQD under vortex beam excitation via a Hertzian dipole array at a fixed irradiation time. 
        (a) Near-field ($z = 2\lambda $) and far-field ($z = 20\lambda$) $\bm{m}_{\rm{GQD}}$ distributions for $l = 1$, $2$, and $3$. 
        (b) Corresponding $x$-component E-field intensity $|E_x|^2$ for $l = 3$ at $z = 2\lambda$ and $z = 20\lambda$.
        }
        \label{figsm:2}
\end{figure*}

As a collective effect induced by topological charge, the anomalously bipolar magnetic moment distribution in GQDs driven by optical vortices is not exclusive to LG beams.  
To demonstrate this generality, we construct a numerically modeled radiative vortex field excited by an array of Hertzian dipole sources in place of an LG beam. 
Specifically, 12 $x$-polarized dipoles are equally spaced on a circle of radius $0.5\lambda$ in the $xy$-source plane ($z = 0$). 
A phase profile $\exp(il\theta)$ relative to the circle's center is imposed on each dipole. 
This configuration generates a radiative vortex beam propagating along the $z$-direction with well-defined topological charge $l$ \cite{Sha2024Spin}.

The E-field of this vortex is governed by the vector Helmholtz equation derived from Maxwell's equations
\begin{align}
\label{eqn:vector_Helmholtz}
        \bm{\nabla} \times \bm{\nabla} \times \bm { E } ( \bm{ R } ) - k ^ { 2 } \bm { E } ( \bm{ R } ) = i \omega \mu_0 \bm { J } ( \bm { R } ),
\end{align}
where $\mu_0$ denotes the vacuum permeability, and $\bm { J } ( \bm { R } )$ represents the source current distribution. 
The resulting E-field is formulated via the dyadic Green's function $\overline{\bf{G}}_e (\bm{R}, \bm{R}^{\prime})$ as
\begin{align}
\label{eqn:E-field_dyadic_Green}
        \bm { E } ( \bm{ R } ) = i \omega \mu_0 \iiint _ { V ^ { \prime } } \overline{{\bf{G}}}_{ e } ( \bm{R} , \bm {R}^{\prime} ) \cdot \bm { J } ( \bm { R } ^ { \prime } ) \diff V ^ { \prime }.
\end{align}

Under this configured optical vortex illumination, the resulting photoinduced magnetic moment distribution $\bm{m}_{\rm{GQD}}$ of the H-GQD is computed using our theoretical framework, consistent with the approach applied for the LG beam. 
We compute the $\bm{m}_{\rm{GQD}}$ distributions at near-field ($z = 2 \lambda$) and far-field ($z = 20\lambda$) observation planes for topological charges $l = 1$, $2$, and $3$, as shown in Fig.~\ref{figsm:2}. 
The corresponding E-field intensity $|E_x|^2$ for $l = 3$ is also shown for comparison.

The absorbed magnetic moment distribution exhibits periodic sign reversals both radially and azimuthally, indicating a spatially modulated IFE.

\bibliography{Supplemental_Mat_Ref}
\bibliographystyle{apsrev4-2}